\documentclass[conference]{IEEEtran}

\usepackage{epsfig,amsmath,amssymb,epsf,amsthm,scalefnt,multirow,subfig}
\usepackage{xcolor}
\usepackage{float}
\usepackage{cite}
\usepackage{psfrag}

\newcommand{\argmax}{\mathop{\mathrm{argmax}}}
\newcommand{\argmin}{\mathop{\mathrm{argmin}}}

\newtheorem{lemma}{Lemma}
\newtheorem*{lemma*}{Lemma}

\newtheorem{corollary}{Corollary}

  \def\cC{{\mathcal{C}}}

 \def\cN{{\mathcal{N}}}  
  \def\cS{{\mathcal{S}}}

\def\argmin{\mathop{\mathrm{argmin}}}
\def\argmax{\mathop{\mathrm{argmax}}}

\def\b0{{\pmb{0}}} 

\def\ba{{\mathbf{a}}} \def\bb{{\mathbf{b}}}  
   \def\bh{{\mathbf{h}}}
   
 \def\bn{{\mathbf{n}}}  
   
  \def\bw{{\mathbf{w}}} \def\bx{{\mathbf{x}}}
\def\by{{\mathbf{y}}}   

\def\bA{{\mathbf{A}}}   
   \def\bH{{\mathbf{H}}}
\def\bI{{\mathbf{I}}}

   \def\bX{{\mathbf{X}}}

\begin{document}

\title{Channel Estimation Techniques for Quantized Distributed Reception in MIMO Systems}
%\title{Distributed Reception for Spatial Multiplexing Systems with Multiple Transmit Antennas}

 %author names and IEEE memberships
\author{Junil Choi$^{\dag}$, David J. Love$^{\dag}$, and D. Richard Brown III$^{\ddag}$\\
\IEEEauthorblockA{$^{\dag}$School of Electrical and Computer
Engineering, Purdue University, West Lafayette, IN\\
$^{\ddag}$Electrical and Computer Engineering Department, Worcester Polytechnic Institute, Worcester, MA}} \maketitle

%\squeezeup\squeezeup\squeezeup\squeezeup
\begin{abstract}
The Internet of Things (IoT) could enable the development of cloud multiple-input multiple-output (MIMO) systems where internet-enabled devices can work as distributed transmission/reception entities.  We expect that spatial multiplexing with distributed reception using cloud MIMO would be a key factor of future wireless communication systems.  In this paper, we first review practical receivers for distributed reception of spatially multiplexed transmit data where the fusion center relies on quantized received signals conveyed from geographically separated receive nodes.  Using the structures of these receivers, we propose practical channel estimation techniques for the block-fading scenario. The proposed channel estimation techniques rely on very simple operations at the received nodes while achieving near-optimal channel estimation performance as the training length becomes large.
\end{abstract}

\section{Introduction}\label{sec1}
The Internet of Things (IoT) could fundamentally change the wireless communication industry as more and more devices (e.g., labtops, smartphones, tablets, and home appliances) are connected through wired/wireless networks \cite{IOT}.  Geographically separated, but closely located, internet-enabled devices could form clusters through a local area network (LAN) and work as massive distributed multiple-input multiple-output (MIMO) systems.  We dub such systems \textit{Cloud MIMO} in this paper.

It is important to point out that cloud MIMO is different from wireless sensor networks (WSNs), i.e., the former is focused on data transmission and reception while the latter is aimed to estimate the behavior of local environments \cite{wsn1,wsn3,dist_detect1}.  Still, there are many similarities between the two, e.g., geographically distributed nodes may cooperate with each other to perform distributed transmission/reception, and it is desirable for each distributed node to perform only simple operations considering processing power or battery life.  These similarities allow us to utilize many techniques developed for WSNs to design cloud MIMO systems.  For example, coded distributed diversity techniques have been proposed to increase the diversity order of distributed reception when the transmitter is equipped with a single antenna \cite{coded_diversity1,coded_diversity2} inspired by exploiting channel coding theory for distributed fault-tolerant classification in WSNs \cite{code_dist_detec1,code_dist_detec2}.

Cloud MIMO will be particularly important at the mobile side.  At base stations, we can deploy a large number of antennas without having strict restriction in space, which is known as massive MIMO \cite{massive_mimo1, massive_mimo3}.  However, it may be difficult to deploy many antennas at a mobile such as a smartphone or a laptop due to its limited space.  The limitation can be overcome by cloud MIMO which exploits the IoT environment.

Recently, the scenario that combines cloud MIMO at the receiver side and spatial multiplexing with multiple transmit antennas is studied in \cite{dist_reception_spatial_multiplex}.  By having only a few quantization bits for the received signal at each receive node, an optimal maximum likelihood (ML) receiver and a suboptimal low-complexity zero-forcing (ZF)-type receiver at the fusion center are proposed.  It is shown analytically and numerically that symbol error rates (SERs) of both receivers can become arbitrary small by increasing the number of distributed receive nodes.  However, the results in \cite{dist_reception_spatial_multiplex} are based on the ideal assumption of perfect global channel knowledge at the fusion center.

In this paper, we extend the work in \cite{dist_reception_spatial_multiplex} and propose practical channel estimation techniques.  Using analytical tools developed in \cite{dist_reception_spatial_multiplex}, we are able to show that channel estimation error can be made arbitrary small by increasing the length of training phase even with small quantization bits at the receive nodes.  Numerical results also show the effectiveness of the proposed channel estimation techniques.

%The rest of the paper is organized as follows.  We define our system model in Section \ref{sec2} and review the ML and ZF-type receivers developed in \cite{dist_reception_spatial_multiplex} in Section \ref{sec3}.  We explain our channel estimation techniques in Section \ref{sec4}.  Simulation results that evaluate the proposed channel estimation techniques are shown in Section \ref{sec5}, and conclusions follow in Section \ref{sec6}.

\textbf{Notation:} Lower and upper boldface symbols represent column vectors and matrices, respectively.  $\|\ba\|$ denotes the two-norm of a vector $\ba$, and $\bA^T$, $\bA^H$, $\bA^\dagger$ are used to denote the transpose, Hermitian transpose, and pseudo inverse of the matrix $\bA$, respectively.  $\mathrm{Re}(\bb)$ and $\mathrm{Im}(\bb)$ denote the real and complex part of a complex vector $\bb$, respectively.  $\mathbf{0}_m$ represents the $m\times 1$ all zero vector, and $\bI_m$ is used for the $m \times m$ identity matrix.  $\mathbb{C}^{m}$ ($\mathbb{R}^{m}$) and $\mathbb{C}^{m\times n}$ ($\mathbb{R}^{m\times n}$) represent the set of all $m \times 1$ complex (real) vectors and the set of all $m \times n$ complex (real) matrices, respectively.

\section{System Model}\label{sec2}

\begin{figure}[t]
  \centering
  %\includegraphics[width=0.8\columnwidth]{conceptual_figure.eps}\\
  \psfrag{h}{$\bH=\begin{bmatrix}\bh_1 & \bh_2 & \cdots \bh_K\end{bmatrix}^H$}
  \psfrag{x}{$\hat{y}_1$}
  \psfrag{y}{$\hat{y}_2$}
  \psfrag{z}{$\hat{y}_K$}
  \includegraphics[width=0.8\columnwidth]{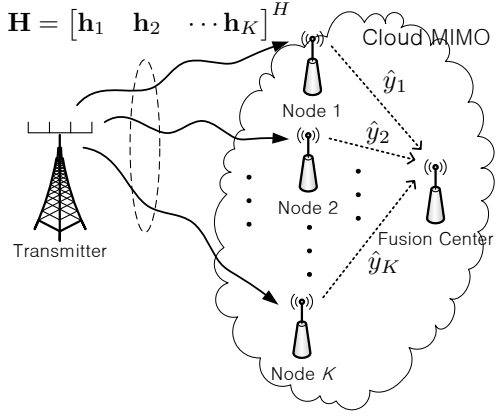}\\
  \caption{The conceptual figure of spatial multiplexing with a cloud MIMO receiver.}\label{concept}
\end{figure}

We consider a network consisting of a transmitter, fusion center, and $K$ geographically separated receive nodes.  We assume the transmitter is equipped with $N_t$ antennas while all other entities in the network have a single antenna.  The transmitter simultaneously transmits $N_t$ independent data symbols by spatial multiplexing, and each receive node conveys its processed (or quantized) received signal to the fusion center through some sort of local area network.  The fusion center decodes the transmitted data symbols using quantized received signals and its (estimated) channel knowledge.  The conceptual figure of this scenario is depicted in Fig. \ref{concept}.

With this setup, the input-output relation is given as\footnote{We consider the block-fading channel model to develop channel estimation techniques in Section \ref{sec4}.}
\begin{equation*}
  \by = \sqrt{\frac{\rho}{N_t}}\bH^H\bx+\bn
\end{equation*}
where
\begin{align*}
\nonumber  \by&=\begin{bmatrix}y_1 & {y}_2 & \cdots & {y}_K\end{bmatrix}^T,\\
\bH&=\begin{bmatrix}\bh_1 & \bh_2 & \cdots & \bh_K\end{bmatrix},\\
\bx&=\begin{bmatrix}x_1& x_2 & \cdots & x_{N_t}\end{bmatrix}^T,\\
\bn&=\begin{bmatrix}n_1 & {n}_2 & \cdots & {n}_K\end{bmatrix}^T.
\end{align*}
Note that $y_k$ is the received signal at the $k$-th received node, $\rho$ is the transmit signal-to-noise ratio (SNR), $\bh_k\sim \cC\cN(\mathbf{0}_{N_t},\bI_{N_t})$ is the independent and identically distributed Rayleigh fading channel vector between the transmitter and the $k$-th receive node, $n_k\sim \cC\cN(0,1)$ is complex additive white Gaussian noise (AWGN) at the $k$-th node, and $x_i$ is the transmitted signal from a standard $M$-ary constellation
\begin{equation*}%\label{constellation}
  \cS = \{s_1,\cdots,s_M\}\subset \mathbb{C}
\end{equation*}
at the $i$-th transmit antenna.  We assume $\cS$ is a phase shift keying (PSK) constellation meaning $|s_m|^2=1$ for all $m$ and $\|\bx\|^2=N_t$.  We assume $x_i$ is drawn from $\cS$ with all symbols equally likely.

Following the same setup as in \cite{dist_reception_spatial_multiplex}, we assume the received signal $y_k$ is quantized with two bits using one bit for each of the real and imaginary parts of $y_k$.  Then the quantized received signal $\hat{y}_k$ is given as
\begin{equation*}%\label{haty_expression}
\hat{y}_k = \mathrm{sgn}(\mathrm{Re}(y_k))+j\mathrm{sgn}(\mathrm{Im}(y_k))
\end{equation*}
where $\mathrm{sgn}(\cdot)$ is the sign function defined as
\begin{equation*}
\mathrm{sgn}(x) = \begin{cases}1& \text{if } x\geq 0 \\
-1 & \text{if } x<0\end{cases}.
\end{equation*}
We assume $\hat{y}_k$ can be forwarded to the fusion center without any error.  The collection of the quantized received signal at the fusion center is given as
\begin{align*}
  \hat{\by}=\begin{bmatrix}\hat{y}_1 & \hat{y}_2 & \cdots & \hat{y}_K\end{bmatrix}^T.
\end{align*}

%\begin{figure}[t]
%  \centering
%  % Requires \usepackage{graphicx}
%  %\includegraphics[width=0.8\columnwidth]{conceptual_figure.eps}\\
%  \psfrag{h}{$\bH=\begin{bmatrix}\bh_1 & \bh_2 & \cdots \bh_K\end{bmatrix}^H$}
%  \psfrag{x}{$\hat{y}_1$}
%  \psfrag{y}{$\hat{y}_2$}
%  \psfrag{z}{$\hat{y}_K$}
%  %one column
%  \includegraphics[width=0.6\columnwidth]{distributed_reception_mimo_psfrag_test.eps}\\
%  % two column
%  %\includegraphics[width=0.8\columnwidth]{distributed_reception_mimo_psfrag_test.eps}\\
%  \caption{The conceptual figure of distributed reception with multiple antennas at the transmitter.  Each receive node is equipped with a single receive antenna.}\label{concept}
%\end{figure}

\section{Review of ML and ZF-type Receivers}\label{sec3}
For the scenario of interest, the optimal ML receiver and the low-complexity ZF-type receiver based on the assumption of perfect global channel knowledge at the fusion center are developed in \cite{dist_reception_spatial_multiplex}.  We briefly review these two receivers in this section.

\subsection{ML receiver}\label{sec3_sub1}
To simplify notation, we first convert all expressions into the real domain as
\begin{align*}
\bH_{\mathrm{R},k}&=\begin{bmatrix} \mathrm{Re}(\bh_k) & -\mathrm{Im}(\bh_k) \\ \mathrm{Im}(\bh_k) & \mathrm{Re}(\bh_k)\end{bmatrix}=\begin{bmatrix}\bh_{\mathrm{R},k,1} & \bh_{\mathrm{R},k,2} \end{bmatrix},\\
\bx_{\mathrm{R}}&=\begin{bmatrix} \mathrm{Re}(\bx) \\ \mathrm{Im}(\bx)\end{bmatrix},\\
\bn_{\mathrm{R},k}&=\begin{bmatrix} \mathrm{Re}(n_k) \\ \mathrm{Im}(n_k)\end{bmatrix},\\
\by_{\mathrm{R},k}&=\begin{bmatrix} \mathrm{Re}(y_k) \\ \mathrm{Im}(y_k)\end{bmatrix}=\begin{bmatrix} y_{\mathrm{R},k,1} \\ y_{\mathrm{R},k,2}\end{bmatrix}
\end{align*}
where
\begin{equation*}
  \bh_{\mathrm{R},k,1}=\begin{bmatrix}\mathrm{Re}(\bh_k) \\ \mathrm{Im}(\bh_k)	\end{bmatrix},\quad \bh_{\mathrm{R},k,2}=\begin{bmatrix}-\mathrm{Im}(\bh_k) \\ \mathrm{Re}(\bh_k)	\end{bmatrix}.
\end{equation*}
Then the input-output relation can be rewritten as
\begin{align}\label{data_phase}
\by_{\mathrm{R},k}=\sqrt{\frac{\rho}{N_t}}\bH_{\mathrm{R},k}^T \bx_{\mathrm{R}} + \bn_{\mathrm{R},k}.
\end{align}
The vectorized version of the quantized $\hat{y}_k$ in the real domain is given as
\begin{equation*}
\hat{\by}_{\mathrm{R},k}=\begin{bmatrix} \mathrm{sgn}(\mathrm{Re}(y_k)) \\ \mathrm{sgn}(\mathrm{Im}(y_k))\end{bmatrix}=\begin{bmatrix} \hat{y}_{\mathrm{R},k,1} \\ \hat{y}_{\mathrm{R},k,2}\end{bmatrix}.\label{real_quantized}
\end{equation*}
We also let $\cS_{\mathrm{R}}$ be
\begin{equation*}
  \cS_{\mathrm{R}}=\left\{\begin{bmatrix}\mathrm{Re}(s_1) \\ \mathrm{Im}(s_1)\end{bmatrix},\cdots,\begin{bmatrix}\mathrm{Re}(s_M) \\ \mathrm{Im}(s_M)\end{bmatrix}\right\}.
\end{equation*}

Based on $\hat{\by}_{\mathrm{R},k}$, the fusion center generates the \textit{sign-refined} channel matrix
\begin{equation*}%\label{sign_refined_H}
\widetilde{\bH}_{\mathrm{R},k}=\begin{bmatrix}\widetilde{\bh}_{\mathrm{R},k,1} & \widetilde{\bh}_{\mathrm{R},k,2} \end{bmatrix}
\end{equation*}
where $\widetilde{\bh}_{\mathrm{R},k,i}$ is defined as
\begin{align*}%\label{sign_refine}
  \widetilde{\bh}_{\mathrm{R},k,i}=\hat{y}_{\mathrm{R},k,i}\bh_{\mathrm{R},k,i}.
\end{align*}

With these definitions, the ML receiver is defined in \cite{dist_reception_spatial_multiplex} as
\begin{align}\label{obj_func}
\hat{\bx}_{\mathrm{R,ML}}=\argmax_{\bx_{\mathrm{R}}'\in\cS_{\mathrm{R}}^{N_t}} \prod_{i=1}^{2}\prod_{k=1}^{K}\Phi\left(\sqrt{\frac{2\rho}{N_t}}\widetilde{\bh}_{\mathrm{R},k,i}^T\bx_{\mathrm{R}}'\right)
\end{align}
where $\Phi(t)=\int_{-\infty}^{t}\frac{1}{\sqrt{2\pi}}e^{-\frac{\tau^2}{2}}d\tau$ and $\cS_{\mathrm{R}}^{N_t}$ is the $N_t$-ary Cartesian product set of $\cS_{\mathrm{R}}$.  We can also define the ML estimator by relaxing the constraint $\bx_{\mathrm{R}}'\in\cS_{\mathrm{R}}^{N_t}$ in \eqref{obj_func} as
\begin{align}\label{ml_est}
\check{\bx}_{\mathrm{R,ML}}=\argmax_{\substack{\bx_{\mathrm{R}}'\in\mathbb{R}^{2N_t},\\ \|\bx_{\mathrm{R}}'\|^2=N_t}} \prod_{i=1}^{2}\prod_{k=1}^{K}\Phi\left(\sqrt{\frac{2\rho}{N_t}}\widetilde{\bh}_{\mathrm{R},k,i}^T\bx_{\mathrm{R}}'\right).
\end{align}
Note that the optimization problem in \eqref{ml_est} is not convex because of the norm constraint on $\bx_{\mathrm{R}}'$.

The following lemma, which is derived in \cite{dist_reception_spatial_multiplex}, shows the performance of the ML estimator.
\begin{lemma}\label{CONVERGE_ML}
For arbitrary $\rho>0$, $\check{\bx}_{\mathrm{R,ML}}$ converges to the true transmitted vector $\bx$ in probability, i.e.,
\begin{equation*}
\check{\bx}_{\mathrm{R,ML}} \stackrel{p}{\longrightarrow} \bx_{\mathrm{R}}
\end{equation*}
as $K\rightarrow \infty$.
\end{lemma}
%\begin{IEEEproof}
The lemma can be proved by using the weak law of large numbers and the stochastic dominance theorem.  Please see Lemma 2 in \cite{dist_reception_spatial_multiplex} for details.
%\end{IEEEproof}

\subsection{ZF-type receiver}
%Using the same real domain notation in the previous subsection, the ZF-type estimator is given as
%\begin{equation*}%\label{zf_est}
%  \check{\bx}_{\mathrm{R,ZF}}=\bH_{\mathrm{R,S}}^{\dagger}\hat{\by}_{\mathrm{R}}
%\end{equation*}
%where
%\begin{align*}%\label{H_stack}
%  \bH_{\mathrm{R,S}}=\begin{bmatrix}\bH_{\mathrm{R},1}^T & \bH_{\mathrm{R},2}^T \cdots & \bH_{\mathrm{R},K}^T\end{bmatrix}^T.
%\end{align*}
The low-complexity ZF-type receiver is developed in \cite{dist_reception_spatial_multiplex}, which is given as
\begin{equation*}%\label{zf_est}
  \check{\bx}_{\mathrm{ZF}}=\left(\bH^H\right)^{\dagger}\hat{\by}.
\end{equation*}
Based on $\check{\bx}_{\mathrm{ZF}}$, the fusion center can perform symbol-by-symbol detection as
\begin{equation*}
  \hat{x}_{\mathrm{ZF},i}=\argmin_{x'\in \cS}|\check{x}_{\mathrm{ZF},i}-x'|^2
\end{equation*}
where $\check{x}_{\mathrm{ZF},i}$ is the $i$-th element of $\check{\bx}_{\mathrm{ZF}}$.

To show the performance of the ZF-type estimator, we first define the mean-squared error (MSE) between $\bx_{\mathrm{R}}$ and $\check{\bx}_{\mathrm{R,ZF}}$ as
\begin{equation*}%\label{avg_MSE}
  \mathrm{MSE}_{\mathrm{ZF}} = \frac{1}{N_t}\mathrm{E}\left[\|\bx-\check{\bx}_{\mathrm{ZF}}\|^2\right]
\end{equation*}
where the expectation is taken over the realizations of channel and noise.  With reasonable assumptions, the MSE of the ZF-type estimator is derived in \cite{dist_reception_spatial_multiplex}, which is rewritten in the following lemma.
\begin{lemma}\label{zf_anal}
If we approximate the quantization error using an additional Gaussian noise $\bw$ as
\begin{equation*}%\label{quant_approx}
  \hat{\by}= \sqrt{\frac{\rho}{N_t}}\bH^H\bx + \bn+\bw,
\end{equation*}
with $\bw\sim \cC\cN(\mathbf{0}_K,\sigma^2_q \frac{\rho}{N_t}\bI_K)$ and assume $\frac{1}{K}\bH\bH^H = \mathbf{I}_{N_t}$, the MSE of the ZF-type estimator is given as
\begin{equation*}
  \mathrm{MSE}_{\mathrm{ZF}} =\frac{N_t\rho^{-1}+\sigma_q^2}{K}.
\end{equation*}.
\end{lemma}
%\begin{IEEEproof}
The lemma can be shown using the analytical tools developed in frame expansion \cite{frame_exp1}.  Please see Lemma 4 in \cite{dist_reception_spatial_multiplex} for details.
%\end{IEEEproof}

Note that the assumption $\frac{1}{K}\bH\bH^H = \mathbf{I}_{N_t}$ holds when $K\rightarrow \infty$.  Moreover, Lemma \ref{zf_anal} shows that the MSE of the ZF-type estimator can be made arbitrary small by increasing the number of receive nodes.

\section{Channel Estimation Techniques}\label{sec4}
Note that the analytical results in the previous section are based on the assumption of perfect channel knowledge at the fusion center.  In practice, global channel knowledge should be estimated using training signals.  Moreover, channel estimation techniques should be based on simple operations at the receive nodes as for distributed reception.  Because the channel between the transmitter and each receive node can be estimated separately, we focus on the channel vector of $k$-th receive node $\bh_k$.

To develop channel estimation techniques, we consider a block-fading channel, i.e., the channel is static for the coherence block length of $L$ channel uses and changes independently from block-to-block.  Then the input-output relation at the $k$-th receive node can be rewritten as
\begin{equation*}
  y_{k,m}[\ell] = \sqrt{\frac{\rho}{N_t}}\bh_{k,m}^H\bx_m[\ell]+n_{k,m}[\ell]
\end{equation*}
for the $\ell$-th channel use in the $m$-th fading block.

We assume that the first $T<L$ channel uses are used for a training phase and the remaining $L-T$ channel uses are dedicated to a data communication phase.  We can write the first $T$ received signals into a vector form as
\begin{equation*}
  \by_{k,m,train} = \sqrt{\frac{\rho}{N_t}}\bX_{m,train}^H\bh_{k,m}+\bn_{k,m,train}
\end{equation*}
where
\begin{align*}
    \by_{k,m,train}&=\begin{bmatrix}y_{k,m}[0] & y_{k,m}[1] \cdots & y_{k,m}[T-1]\end{bmatrix}^H,\\
    \bX_{m,train}&=\begin{bmatrix}\bx_m[0] & \bx_m[1] \cdots & \bx_m[T-1]\end{bmatrix},\\
    \bn_{k,m,train}&=\begin{bmatrix}n_{k,m}[0] & n_{k,m}[1] \cdots & n_{k,m}[T-1]\end{bmatrix}^H.\\
\end{align*}
Note that $\by_{k,m,train}\in \mathbb{C}^{T}$, $\bX_{m,train}\in \mathbb{C}^{N_t\times T}$, and $\bn_{k,m,train}\in \mathbb{C}^{T}$.  In the training phase, $\bX_{m,train}$ is known to both the transmitter and the fusion center while $\bh_{k,m}$ needs to be estimated at the fusion center.  We focus on unitary training and assume $\bX_{m,train}$ satisfies \cite{training4}
\begin{align*}
\bX_{m,train}^H \bX_{m,train} =\bI_{T}\quad &\text{if }N_t\geq T, \\
\bX_{m,train} \bX_{m,train}^H =\frac{T}{N_t} \bI_{N_t}\quad &\text{if }N_t<T.
\end{align*}
The normalization term $T/N_t$ in the case of $N_t<T$ ensures the average transmit SNR is equal to $\rho$ in each channel use.

Similar to Section \ref{sec3_sub1}, we can reformulate these expressions into the real domain as
\begin{equation}\label{train_phase}
  \by_{\mathrm{R},k,m,train} = \sqrt{\frac{\rho}{N_t}}\bX_{\mathrm{R},m,train}^T\bh_{\mathrm{R},k,m}+\bn_{\mathrm{R},k,m,train}
\end{equation}
where
\begin{align*}
\by_{\mathrm{R},k,m,train}&=\begin{bmatrix}\mathrm{Re}\left(\by_{k,m,train}\right) \\ \mathrm{Im}\left(\by_{k,m,train}\right)\end{bmatrix},\\
\bX_{\mathrm{R},m,train}&=\begin{bmatrix} \mathrm{Re}(\bX_{m,train}) & -\mathrm{Im}(\bX_{m,train}) \\ \mathrm{Im}(\bX_{m,train}) & \mathrm{Re}(\bX_{m,train})\end{bmatrix},\\
\bh_{\mathrm{R},k,m}&=\begin{bmatrix}\mathrm{Re}\left(\bh_{k,m}\right) \\ \mathrm{Im}\left(\bh_{k,m}\right)\end{bmatrix},\\
\bn_{\mathrm{R},k,m,train}&=\begin{bmatrix}\mathrm{Re}\left(\bn_{k,m,train}\right) \\ \mathrm{Im}\left(\bn_{k,m,train}\right)\end{bmatrix}.
\end{align*}
It is easy to show that $\by_{\mathrm{R},k,m,train}\in\mathbb{R}^{2T}$, $\bX_{\mathrm{R},m,train}\in\mathbb{R}^{2N_t \times 2T}$, $\bh_{\mathrm{R},k,m}\in\mathbb{R}^{2N_t}$, and $\bn_{\mathrm{R},k,m,train}\in\mathbb{R}^{2T}$.

It is important to point out that \eqref{train_phase} has the same form as \eqref{data_phase} while the roles of the channel and the training signal are reversed.  Thus, using the same techniques in Section \ref{sec3}, we can develop channel estimators based on the knowledge of the quantized signal $\hat{\by}_{\mathrm{R},k,m,train}$ and $\bX_{\mathrm{R},m,train}$.

We define the $i$-th column of $\bX_{\mathrm{R},m,train}$ as $\bx_{\mathrm{R},m,train,i}$ and
\begin{equation*}
\hat{y}_{\mathrm{R},k,m,train,i} = \mathrm{sgn}(y_{\mathrm{R},k,m,train,i})
\end{equation*}
where $y_{\mathrm{R},k,m,train,i}$ is the $i$-th element of $\by_{\mathrm{R},k,m,train}$.  Then the sign-refinement based on $\hat{y}_{\mathrm{R},k,m,train,i}$ is performed as
\begin{equation*}
\widetilde{\bx}_{\mathrm{R},m,train,i}=\hat{y}_{\mathrm{R},k,m,train,i}\bx_{\mathrm{R},m,train,i},
\end{equation*}
and the ML channel estimator is given as
\begin{align}
\nonumber \check{\bh}_{\mathrm{R},k,m,\mathrm{ML}}&=\argmax_{\bh_{\mathrm{R}}'\in\mathbb{R}^{2N_t}} \prod_{i=1}^{2T}\Phi\left(\sqrt{\frac{2\rho}{N_t}}\widetilde{\bx}_{\mathrm{R},m,train,i}^T\bh_{\mathrm{R}}'\right)\\
&=\argmax_{\bh_{\mathrm{R}}'\in\mathbb{R}^{2N_t}} \sum_{i=1}^{2T}\log\left(\Phi\left(\sqrt{\frac{2\rho}{N_t}}\widetilde{\bx}_{\mathrm{R},m,train,i}^T\bh_{\mathrm{R}}'\right)\right).\label{ml_ch_est}
\end{align}
Because $\Phi(\cdot)$ is a log-concave function, we can efficiently solve \eqref{ml_ch_est} using standard convex optimization methods \cite{convex_opt}.  However, if $T$ is not large enough, the ML channel estimator returns an inaccurate channel estimate because there are not enough samples to estimate the true channel.  For example, if we only consider $i=1$, then there are many possible choices for $\bh_{\mathrm{R}}'$ that give $\Phi(\cdot)$ equals to one.  This trend is shown by numerical studies in Section \ref{sec5}.

We can also define the ZF-type channel estimator as
\begin{equation}\label{zf_ch_est}
\check{\bh}_{\mathrm{R},k,m,\mathrm{ZF}}=\left(\bX_{\mathrm{R},m,train}^T\right)^{\dagger}\hat{\by}_{\mathrm{R},k,m,train}.
\end{equation}
If $N_t<T$, then \eqref{zf_ch_est} can be also written as
\begin{equation*}
\check{\bh}_{\mathrm{R},k,m,\mathrm{ZF}}=\frac{N_t}{T}\bX_{\mathrm{R},m,train}\hat{\by}_{\mathrm{R},k,m,train}
\end{equation*}
because $\bX_{m,train} \bX_{m,train}^H =\frac{T}{N_t} \bI_{N_t}$.  Note that the norm of $\check{\bh}_{\mathrm{R},k,m,\mathrm{ZF}}$ highly depends on the norm of $\hat{\by}_{\mathrm{R},k,m,train}$.  However, $\hat{\by}_{\mathrm{R},k,m,train}$ is based on the sign function and does not have any norm information of $\by_{\mathrm{R},k,m,train}$.  Thus, we consider the MSE of the normalized channel estimate, which is defined as
\begin{equation*}
  \mathrm{MSE}_{x,\bh}=\frac{1}{N_t}\mathrm{E}\left[\left\Vert\frac{\bh_{\mathrm{R},k,m}}{\|\bh_{\mathrm{R},k,m}\|}-\frac{\check{\bh}_{\mathrm{R},k,m,x}}{\|\check{\bh}_{\mathrm{R},k,m,x}\|}\right\Vert^2\right],
\end{equation*}
for the performance metric of a receiver $x$ in Section \ref{sec5}.

Although similar, we are not able to apply Lemma \ref{CONVERGE_ML} to analyze the performance of the ML channel estimator because Lemma \ref{CONVERGE_ML} is based on the norm constraint on $\bx_{\mathrm{R}}'$ while the ML channel estimator does not have such a constraint.  However, we can still analyze the performance of the ZF-type channel estimator with the same assumption of quantization error as in Lemma \ref{zf_anal}.

\begin{corollary}\label{cor_1}
If $N_t<T$ and we approximate the quantization error of the first $T$ received training signals in the $m$-th fading channel at the $k$-th receive node using an additional Gaussian noise $\bw_{k,m,train}$ as
\begin{align*}
  \hat{\by}_{k,m,train}= \sqrt{\frac{\rho}{N_t}}\bX^T_{m,train}\bh_{k,m}+\bn_{k,m,train}+\bw_{k,m,train}
\end{align*}
with $\bw_{k,m,train}\sim \cC\cN(\mathbf{0}_{T},\sigma^2_{q,train} \frac{\rho}{N_t}\bI_{T})$, the MSE of the ZF-type channel estimator is given as
\begin{equation*}
  \mathrm{MSE}_{\mathrm{ZF},train}=\frac{N_t^3\rho^{-1}+N_t^2\sigma_{q,train}^2}{T}.
\end{equation*}
\end{corollary}
The result is a direct consequence of Lemma \ref{zf_anal}.  The lemma shows that we can make the MSE of the ZF-type channel estimator arbitrary small by increasing the length of the training phase $T$.  Numerical studies in Section \ref{sec5} shows the same result holds for the ML channel estimator as well.

\section{Simulation Results}\label{sec5}
%\begin{figure*}
%\centering
%\subfloat[Moderate training length $T$.]{
%% one column
%\includegraphics[width=0.9\columnwidth]{MSE_1.eps}
%% two column
%%\includegraphics[width=1\columnwidth]{MSE_fixedSNR10_log.eps}
%\label{mse1}
%}
%\subfloat[Large training length $T$.]{
%% one column
%\includegraphics[width=0.9\columnwidth]{MSE_2.eps}
%% two column
%%\includegraphics[width=1\columnwidth]{MSE_fixedK50_log.eps}
%\label{mse2}
%}
%\caption{The MSEs of the normalized channel estimates with $N_t=4$ and different values of $T$ (training channel uses) and $\rho$ (transmit SNR).}
%\label{mse_fig}
%\end{figure*}
\begin{figure}[t]
  \centering
  \includegraphics[width=1\columnwidth]{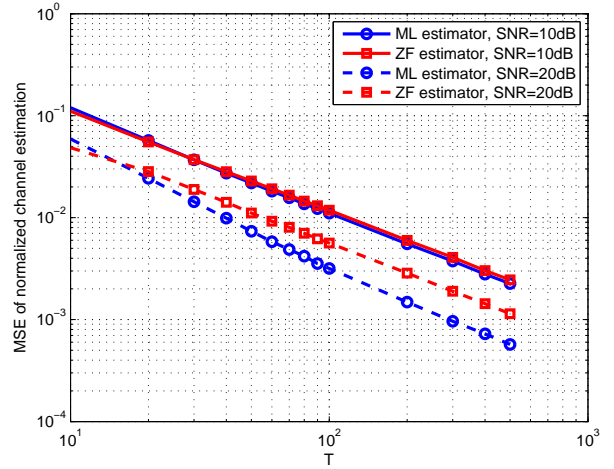}\\
  \caption{The MSEs of the normalized channel estimates with $N_t=4$ and different values of $T$ (training channel uses) and $\rho$ (transmit SNR).}\label{mse_fig}
\end{figure}
We perform Monte-Carlo simulations to evaluate the proposed channel estimation techniques.  In Fig. \ref{mse_fig}, we first compare the MSEs of the normalized channel estimates of the ML and ZF-type channel estimators, i.e., $\mathrm{MSE}_{\mathrm{ML},\bh}$ and $\mathrm{MSE}_{\mathrm{ZF},\bh}$, defined in the previous section.  The results are averaged over 10000 channel realizations of a single receive node with $N_t=4$.  As expected, the MSEs of both channel estimators go to zero as $T$ increases.  The convergence rate of the ZF-type channel estimator (and the ML channel estimator as well) is proportional to $\frac{1}{T}$ as derived in Corollary~\ref{cor_1}.  Note that the ML channel estimator is inferior to the ZF-type channel estimator when $T$ is small, which is explained in Section \ref{sec4}.  However, the ML channel estimator outperforms the ZF-type channel estimator as $T$ becomes large.  The gap between the two channel estimators is not significant with 10dB SNR but there is a notable gap between the two with 20dB SNR.

\begin{figure}[t]
  \centering
  \includegraphics[width=1\columnwidth]{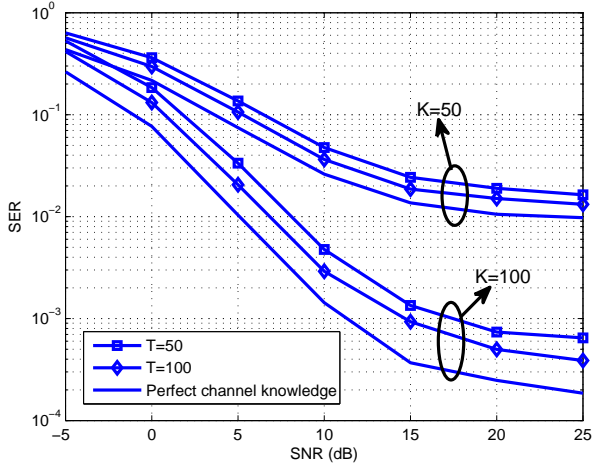}\\
  \caption{Symbol error rate (SER) vs. SNR in dB scale for the ZF-type receiver with different levels of channel estimation quality. We set $N_t=4$ and 8PSK for $\cS$.}\label{SER}
\end{figure}

In Fig. \ref{SER}, we plot the SER of the ZF-type receiver\footnote{Because we consider the normalized channel estimates, we do not compare the SER of the ML receiver.} based on the ZF-type channel estimator.  The SER is defined as 
\begin{equation*}
  \mathrm{SER} = \frac{1}{N_t}\sum_{n=1}^{N_t}\mathrm{E}\left[\mathrm{Pr}\left(\hat{x}_n\neq x_n\mid \bx~\text{sent},\bH,\bn,\rho,N_t,K,\cS \right)\right]
\end{equation*}
where the expectation is taken over $\bx$, $\bH$, and $\bn$.  We again fix $N_t=4$ and adopt 8PSK constellation for $\cS$.  As $T$ increases, the SER performance approaches to the case of perfect channel knowledge.  Although the ZF-type receiver suffers from an error rate floor in high SNR regime, the error floor can be mitigated by having more receive nodes for both cases of perfect and estimated channel knowledge.

\section{Conclusion}\label{sec6}
We studied the scenario that combines spatial multiplexing and cloud MIMO for distributed reception in this paper. To relax the ideal assumption of perfect global channel knowledge considered in \cite{dist_reception_spatial_multiplex}, we proposed practical channel estimation techniques that rely on very simple operations at the receive nodes.  We showed that even with very coarse quantization at the receive nodes, the fusion center can estimate the channel with high accuracy if the length of the training phase is sufficiently large.  To reduce the overhead of the training phase, we can exploit long-term channel statistics, i.e., spatial and/or temporal correlation of channels as in \cite{cl_training_jstsp}, which would be an interesting future research topic.

\bibliographystyle{IEEEtran}
\bibliography{refs}

\end{document}